\documentclass[%
 reprint,
 amsmath,amssymb,
 aps,
noeprint
]{revtex4-2}

\usepackage{graphicx}
\usepackage{dcolumn}
\usepackage{bm}
\usepackage{amsmath}
\usepackage{longtable}

\begin{document}

\preprint{APS/123-QED}

\title{Impact of substrates and quantum effects on exciton line shapes of 2D semiconductors at room temperature}

\author{Jorik van de Groep$^{1,2}$}
\author{Qitong Li$^{1}$}
\author{Jung-Hwan Song$^{1}$}%
\author{Pieter G. Kik$^{3}$}%
\author{Mark L. Brongersma$^{1}$}%

\affiliation{%
 $^{1}$Geballe Laboratory for Advanced Materials, Stanford University \\
 Stanford, CA 94305, USA 
}%
\affiliation{%
 $^{2}$Van der Waals-Zeeman Institute, Institute of Physics, University of Amsterdam \\
 Amsterdam, 1098 XH, The Netherlands 
}%
\affiliation{%
 $^{3}$ CREOL, The College of Optics and Photonics, University of Central Florida\\ Orlando, FL 32816, USA
}

\date{\today}

\begin{abstract}
Exciton resonances in monolayer transition-metal dichalcogenides (TMDs) provide exceptionally  strong light-matter interaction at room temperature. Their spectral line shape is critical in the design of a myriad of optoelectronic devices, ranging from solar cells to quantum information processing. However, disorder resulting from static inhomogeneities and dynamical fluctuations can significantly impact the line shape. Many recent works experimentally evaluate the optical properties of TMD monolayers placed on a substrate and the line shape is typically linked directly to the material's quality. Here, we highlight that the interference of the substrate and TMD reflections can strongly influence the line shape. We further show how basic, room-temperature reflection measurement allow investigation of the quantum mechanical exciton dynamics by systematically controlling the substrate reflection with index-matching oils. By removing the substrate contribution with a properly chosen oil, we can extract the excitonic decay rates including the quantum mechanical dephasing rate. The results provide valuable guidance for the engineering of exciton line shapes in layered nanophotonic systems.
\end{abstract}

\maketitle


\section{\label{sec:into}Introduction}
 The strong light-matter interaction offered by excitons in monolayer 
 TMDs has generated wide spread interest, with potential applications in atomically-thin transistors \cite{Radisavljevic2011,Daus2021}, photodetectors \cite{Yin2012,Lopez-Sanchez2013}, solar cells \cite{NassiriNazif2021,NassiriNazif2021a}, light sources \cite{Ross2013,Frisenda2018}, and quantum optics devices \cite{Srivastava2015,Palacios-Berraquero2016}. Unlike those in bulk semiconductors, excitons in monolayer TMDs are quantum confined and exhibit large binding energies due to strongly reduced dielectric screening \cite{Chernikov2014}. Exciton binding energy of 100's of meV are quite common and render highly stable excitons, even at room temperature \cite{Ugeda2014}. The spectrally-sharp absorption features associated with excitons strongly impact the material's dielectric constant and can dominate the optical response in the visible spectral range \cite{Cao2015,Mak2016,Mueller2018}. The exciton properties are highly sensitive to temperature \cite{Ross2013,Epstein2020}, external fields \cite{Stier2016,Stier2018}, carrier density \cite{Ross2013,Newaz2013,Yu2017}, and strain \cite{Lloyd2016,Aslan2018}. Combined with the facile integration of monolayer materials in complex nanophotonic systems \cite{Krasnok2018,Zhang2020}, the high tunability of the exciton can be leveraged to realize a variety of dynamic nanophotonic devices \cite{Ross2014,Scuri2018,Back2018,Epstein2020}.
 
For applications in light modulation and wavefront shaping \cite{Scuri2018,Back2018,VandeGroep2020,Datta2020,Li2020,Biswas2021}, engineering of the exciton line shape is essential. It has been shown that TMD monolayers can serve as atomically-thin mirrors with a well-defined susceptibility. At cryogenic temperatures, the susceptibility is often dominated by the exciton resonances and their behavior is captured by a Lorentzian resonance. At room temperature, the resonances are less pronounced and the spectrally-broad background from other contributions are often of a similar strength. The experimentally observed line shape of TMD monolayers varies with the material quality and the dielectric environment of the monolayer. Previous works have carefully studied the optical properties of suspended monolayers and monolayers on substrates - both using reflection measurements on small exfoliated flakes \cite{Li2014} and ellipsometric measurements on large-area monolayers grown through chemical-vapor deposition \cite{Park2014,Liu2014}. The results give insight into the material's optical response and can be used in 2D and 3D theoretical models \cite{Li2018}. However, systematic investigation of the interference of the monolayer reflection and the substrate reflection is essential for understanding the spectral line-shapes observed in experiments \cite{Scuri2018,Back2018,Epstein2020a}. 

The reflectance of a TMD monolayer placed on a substrate cannot be treated classically, but also includes the quantum mechanical nature of the radiative and non-radiative decay of the excitons as well as their dephasing behavior. Recent studies showed that temperature-dependent, non-radiative relaxation and pure dephasing rates can be extracted from asymmetric reflection spectra \cite{Scuri2018,Epstein2020}. The asymmetric line-shape in these works originates from interference between the optically-driven, spectrally-sharp exciton radiation and a broad-band cavity reflection. This interference is sensitive to both the amplitude and phase of the exciton re-radiation - which enables one to distinguish between non-radiative decay and pure dephasing. Like classical interference, a quantum mechanical analysis involves radiative coupling to the far-field that is altered by the presence of a substrate \cite{Epstein2020}. As the exciton resonances at room-temperature are weaker than at cryogenic temperatures, the reflectance from the substrate and monolayers can be similar in magnitude and this can lead to strong interference effects without the presence of a cavity. The significant impact of the environment on the exciton line shape makes it of value to develop a detailed understanding of the constituent effects that govern the spectral line shape of strong exciton resonances in layered nanophotonic systems.

Here, we perform reflectance measurements on TMD monolayers placed on substrates and control the substrate contribution systematically using oils with varying refractive index. We use commercial large-area monolayer WS$_{2}$ on a fused silica substrate and determine the optical constants using spectroscopic ellipsometry. Using these constants we employ a total-field scattered-field (TFSF) analysis to isolate the monolayer contribution to the measured reflectivity and gain an understanding of the interference between the monolayer and substrate reflections. By isolating the monolayer signal with an index-matched oil, we then demonstrate that the quantum mechanical exciton properties can be retrieved - even at room temperature. This is enabled by the highly asymmetric spectral line shape resulting from the interference between the background dielectric permittivity (all other transitions) and the excitonic contribution to the permittivity.  This analysis thus does not rely on interference with the substrate or external cavity. These results provide guidelines to design and leverage the substrate contribution in more complex nanophotonic geometries and tunable optical elements.

\section{\label{sec:Experiment}Experiment}
\begin{figure*}
\includegraphics[width=17cm]{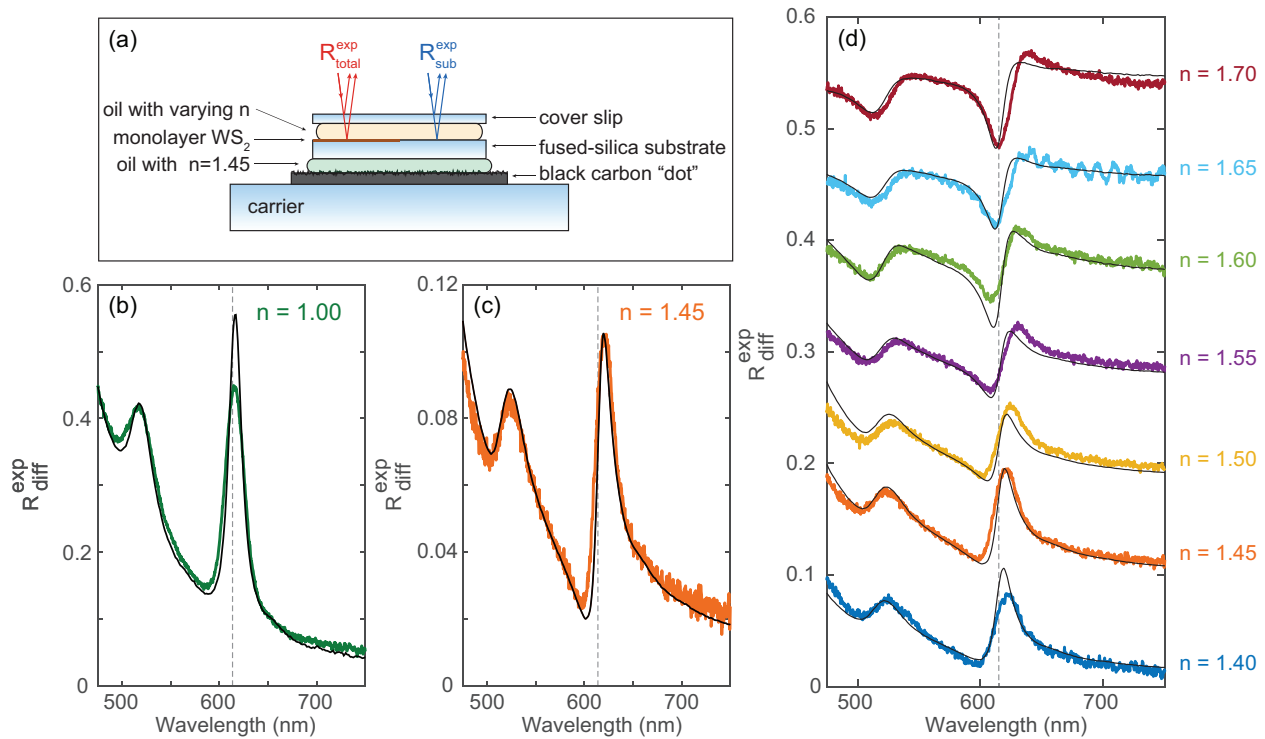}
\caption{\label{fig1}(a) Schematic of the experimental geometry (for $n>1$) to measure the reflection of the bare (right, $R_{sub}^{exp}$) and WS$_{2}$-covered substrate (left, $R_{total}^{exp}$). The reflection from the backside of the substrate is suppressed by index-matching oil ($n=1.45$) and a black carbon dot. (b,c) Measured (color) and calculated (black) differential reflection for monolayer WS$_{2}$ in air (b) and embedded in oil with $n=1.45$ (c). The calculated spectra are based on transfer-matrix calculations with measured ellipsometry data as an input. (d) Measured (color) and calculated (black) differential reflection for oils with $n=1.40-1.70$. Spectra are offset for visibility. The vertical dashed line in (b-d) indicates $\lambda=615$~nm, corresponding to the peak in exciton absorption.}
\end{figure*}

To experimentally control the substrate contribution to the measured reflection spectrum, we employ a series of refractive index oils in the range $n = 1.40-1.70$ (Cargille labs) in combination with a glass cover slip and carrier substrate (Fig.~\ref{fig1}a). We obtain large-area monolayer WS$_{2}$ (1$\times$1~cm$^{2}$) on a fused-silica (FS) substrate commercially (2D Semiconductors). To realize an internal reference measurement on the bare substrate, we etch away the monolayer WS$_{2}$ on half of the surface area using physical masking and a reactive-ion etching process (Ar ion milling). To isolate the reflection of the oil-WS$_{2}$-FS interface, the reflection from the backside of the substrate needs to be suppressed. We use a black ``carbon dot" mounted on a glass carrier substrate to function as a broadband absorber and employ index-matching oil with $n=1.45$ between the FS substrate and carbon dot to prevent reflections from the FS-air interface.  We measure the reflection of the WS$_{2}$-covered FS ($R_{total}^{exp}$, red in Fig.~\ref{fig1}a) with air ($n=1.00$) as a superstrate and after depositing an oil with varying $n = 1.40-1.70$ on top, covered by a silica cover slip. To remove the reflection contributions from the top cover slip, we also collect a reference measurement of the bare FS ($R_{sub}^{exp}$, blue in Fig.~\ref{fig1}a) for each choice of oil, and the relative differential reflection $R_{diff}^{exp}=\frac{R_{total}^{exp}-R_{sub}^{exp}}{R_{sub}^{exp}}=\frac{\Delta R^{exp}}{R_{sub}^{exp}}$ is determined.  We perform the reflection measurements using a Nikon C2 confocal microscope, equipped with a halogen light source, 20$\times$ microscope objective (Nikon, 3.8 mm working distance, 0.4 NA) and 90~$\mu$m pinhole. The aperture stop of the K\"ohler-illumination is set to its lowest setting to minimize the angular spread of the incident light. The reflection spectra are recorded using a Princeton Instruments grating spectrograph (150 lines/mm) and PIXIS camera. Each reflection spectrum is the average of 25 recordings (6~s integration each) and the dark spectrum is subtracted. We use local defects in the monolayer and substrate as spatial markers to measure the reflectance spectra at exactly the same sample area for each refractive index oil.

Figure~\ref{fig1}b shows the reflection spectrum of the monolayer on FS in air. A clear peak is observed around $\lambda=615$~nm that corresponds to the A-exciton in WS$_{2}$~\cite{Li2014}; the second peak around $\lambda=517$~nm corresponding to the B-exciton~\cite{Li2014} is also well-pronounced. The line shape of the A-exciton is nearly symmetric and appears close to $\lambda=615$~nm (grey dashed line), corresponding to the peak in excitonic absorption (retrieved from ellipsometry, discussed below), suggesting that the reflection spectrum of WS$_{2}$ on a substrate is dominated by the imaginary component ($\varepsilon_{2}$) of the dielectric constant $\varepsilon=\varepsilon_{1}+i \varepsilon_{2}$ \cite{Li2014}. The exciton line shape becomes strongly asymmetrical when the refractive index of the superstrate is changed to $n=1.45$ (Fig.~\ref{fig1}c) - matched to the index of the FS substrate. Under these index-matched conditions, the reflection signal is minimized to the ``pure" monolayer scattering without the interference with a substrate reflection. The asymmetrical line shape does not follow $\varepsilon_{2}$, and the reflection peak is shifted with respect to $\lambda=615$~nm. The large differences between Fig.~\ref{fig1}b,c demonstrate that the experimentally-observed exciton line shape is strongly influenced by the presence of the substrate. We will show that all observed changes in the nontrivial line shape can be understood in terms of the basic interference of the monolayer reflection and the substrate reflection. Dielectric screening of the exciton by the oil~\cite{Lin2014,Raja2017} does not result into significant shifts in the exciton binding energy that cannot be explained by the interference.

To explore the strong changes in the line shape in more detail, Fig.~\ref{fig1}d shows the reflection spectra for oils with refractive indices ranging from $n=1.40 - 1.70$, slowly increasing in steps of $0.05$. Each spectrum shows a strong excitonic response around $\lambda=615$~nm, while the line shape changes significantly as a function of refractive index. The line shape changes from a peak around $\lambda=615$~nm ($n=1.40$), to an asymmetric Fano line shape ($n=1.55$), and finally to a dip around $\lambda=615$~nm ($n=1.70$). Note that the high-frequency oscillations in the spectrum for $n=1.65$ originate from coherent  interference fringes in the thin oil layer due to the more pronounced spreading as a result of capillary forces of this particular oil. The gradual transition in the oil refractive index slowly changes the amplitude and phase of the substrate contribution to the reflection signal. For $n<1.45$ there is a small substrate reflection with a $\pi$ radian phase shift. For $n=1.45$ there is practically no substrate contribution, and for $n>1.45$ there is an increasing substrate reflection with no phase shift. As such, these results enable us to systematically analyze the interference of the monolayer scattering and the substrate reflection.

\begin{figure*}
\includegraphics[width=17cm]{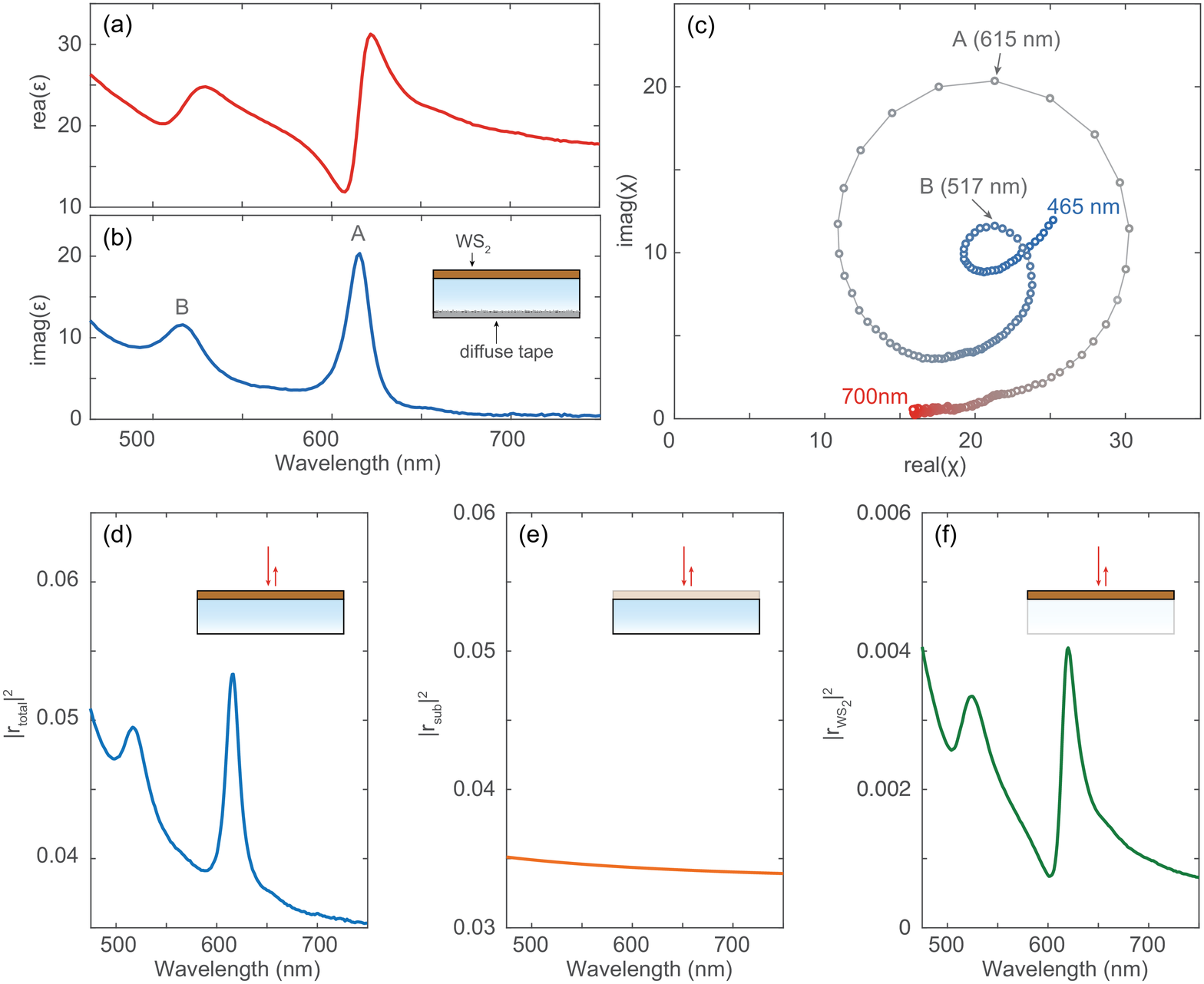}
\caption{\label{fig2}(a,b) Real (a) and imaginary (b) part of the dielectric constant retrieved from ellipsometric measurements on large-area monolayer WS$_{2}$ on a fused-silica substrate. (c) Corresponding complex susceptibility $\chi=\varepsilon-1$ showing the large oscillations in the materials optical response due to the exciton resonances. The color of the plot marks indicates the corresponding wavelength, ranging from 465~nm (blue) to 750~nm (red). The A and B excitons are indicated in (b,c). (d-f) Total-field scattered-field analysis of monolayer WS$_{2}$ scattering on a fused silica substrate in air ($n=1$). The complex reflection coefficient of the bare substrate $r_{sub}$ (e) is subtracted from that of the complete structure $r_{total}$ (d), to yield the complex monolayer reflection coefficient $r_{WS_{2}}$ (f). Despite the single peak observed in the total reflection spectrum (d), the monolayer contribution is asymmetric (f) analogous to Fig.~\ref{fig1}c. The spectra in panels (d-f) are based on transfer-matrix calculations with ellipsometry data as an input.}
\end{figure*}

\section{\label{sec:Model}Model}
To gain better understanding of the monolayer-substrate interference, we use a total-field scattered-field analysis based on transfer-matrix calculations to isolate the monolayer scattering contribution for each oil index, in three steps. 

\subsection{\label{sec:ellipsometry}Spectroscopic Ellipsometry}
First, we perform spectroscopic ellipsometry (Woollam M2000) on the bare FS and monolayer WS$_{2}$ on FS to obtain accurate complex optical constants (dispersive dielectric function $\varepsilon$) for the FS and WS$_{2}$. To remove contributions from the back of the transparent substrate, we apply diffusive tape on the backside (see inset Fig.~\ref{fig2}b). The optical properties of the FS substrate are modeled by two Sellmeier oscillators \cite{Sellmeier1872} outside the spectral range of interest, yielding a refractive index $n_{sub}=1.45$ with a very small dispersion. The WS$_{2}$ is modeled as an isotropic layer with a Tauc-Lorentz oscillator \cite{Jellison1996,Jellison1996a} for the interband transitions and two Lorentz oscillators for the A and B excitons. While the WS$_{2}$ dielectric permittivity is anisotropic by nature (i.e. out-of-plane component is different from in-plane components), we were unable to quantify this anisotropy in the ellipsometric measurements. Despite this, we find that the measured isotropic dielectric function is accurate enough to describe the main features in the quasi-normal incidence reflectance spectra. The monolayer thickness is assumed to be $6.18$~{\AA}, equal to the interlayer spacing in bulk WS$_{2}$ \cite{Wilson1969}. The dispersion of the measured real ($\varepsilon_{1}$) and imaginary ($\varepsilon_{2}$) part of the WS$_{2}$ dielectric constant are shown in Fig.~\ref{fig2}a,b, respectively. The largest absorption is observed for a wavelength of $615$~nm, which denotes the exciton resonance wavelength. While the WS$_{2}$ absorption ($\varepsilon_{2}$) shows a strong and approximately symmetric peak around $\lambda=615$~nm, the real part shows a large asymmetrical oscillation. See Supporting Information section~\ref{SI-ellips} for tabulated data of the measured refractive index $\tilde{n}=n+i\kappa=\sqrt{\varepsilon}$.  

Based on the measured $\varepsilon$, we can already interpret the asymmetrical line shape of the pure monolayer scattering (Fig.~\ref{fig1}c), by realizing that the locally generated scattered fields are proportional to the polarization $\bf{P}=\varepsilon_{0}\chi \bf{E_{i}}$ of the WS$_{2}$. Here, $\varepsilon_{0}$ is the free-space permittivity, $\bf{E_{i}}$ is the local driving electric field, and $\chi=(\varepsilon -1)$ is the complex electrical susceptibility of WS$_{2}$. The spectral dispersion of $\chi$ thus dictates the amplitude and phase of the light scattered by the monolayer, with the strongest scattering when $|\chi|$ is largest. Fig.~\ref{fig2}c shows the measured susceptibility in the complex plane. The exciton resonances give rise to large oscillations on top of the background susceptibility, thereby introducing the asymmetrical line shape in $|\chi|$.

\subsection{\label{sec:TMM}Transfer-Matrix model}
Second, we employ the measured optical constants in a transfer-matrix model (TMM) to calculate the differential reflection for each refractive index value. In the model, we consider only the oil-WS$_{2}$-FS interfaces and assume both the oil and the FS to be semi-infinite. This is valid because there is no correlation between the reflection from the top cover slip and from the oil-WS$_{2}$-FS interfaces that can modify the line shape of the measured reflection signal. The modeled differential reflection $R_{diff}$ was scaled by a constant factor $c$ to match the measured differential reflection $R_{diff}^{exp}$ to account for the fact that the model does not include the relatively large but constant reflection contribution from the cover slip:
\begin{multline}
R_{diff}^{exp}=\frac{R_{total}^{exp}-R_{sub}^{exp}}{R_{sub}^{exp}}\\ 
\approx \frac{R_{top}+ T_{top}^{2}R_{total}- R_{top}-T_{top}^{2}R_{sub}}{ R_{top}+T_{top}^{2}R_{sub}}\\
\approx c\frac{ R_{total}-R_{sub}}{R_{sub}}=c R_{diff}.
\end{multline}
$R_{total}$ and $R_{sub}$ represent the modeled reflection from the oil-WS$_{2}$-FS interfaces and bare oil-FS interface, respectively. $R_{top}$ and $T_{top}$ denote the reflectance and transmittance of the top cover slip. The resulting differential reflection spectra are shown in Fig.~\ref{fig1}b-d as solid black lines, and show good agreement with the measured spectra. Interestingly, the agreement for $n=1.45$ with no substrate contribution is very good (Fig.~\ref{fig1}c), except for a small discrepancy around $\lambda=600$~nm. For all other refractive indices the calculated exciton line shape exhibits a slightly smaller line width and larger amplitude than the measured spectra. This is especially remarkable as the $\varepsilon$ used in the TMM calculations is based on a large-area (mm-scale) ellipsometry measurement that could exhibit significant inhomogeneous broadening. The reflection measurements on the other hand use a collection area with a diameter of $\sim$4.5~$\mu$m, and thereby have a lower sensitivity to non-uniformity in the WS$_{2}$ material. We attribute the broadening in the measured spectra to a slight sample degradation as a result of the sample processing steps that were performed after the ellipsometry measurements, but before the reflection measurements. Our analysis assumes negligible inhomogeneous broadening. Despite the small deviations in line width, the TMM calculated spectra show good agreement with the measured spectra, validating the use of TMM to study the monolayer-substrate interference.

\subsection{\label{sec:TFSF}Total-Field Scattered-Field analysis}
Third, to gain further insight into the observed changes in line shape, we use the TMM to perform a total-field scattered-field (TFSF) analysis that allows us to isolate the monolayer reflection amplitude and phase. Given the fact that the wavelength of the incident light is much larger than the thickness of the monolayer WS$_{2}$, the reflection contribution from the 2D layer can be modeled as the radiation from an equivalent surface current driven by the local electric field at the oil-FS interface. Therefore, The total complex field reflection results from the superposition of the reflection field from the oil-FS interface and the radiation field from the equivalent surface current $r_{total}= r_{sub}+ r_{WS_{2}}$. This allows us to expand the modeled differential reflection $R_{diff}$ as:
\begin{equation}
R_{diff}=\frac{|r_{WS_{2}}|^{2}+2 \Re{\left({r_{WS_{2}}}^{*}r_{sub}\right)}}{|r_{sub}|^{2}}.
\end{equation}
The above equation verifies that the measured line shape is directly related to the interference between the reflection from the substrate and the reflection from the 2D layer. We first calculate the complex field reflection coefficient of the substrate \textit{with} TMD layer $r_{total}$ (inset of Fig~\ref{fig2}d). $|r_{total}|^{2}$ corresponds to what is typically measured in a reflection experiment, and is plotted in Fig.~\ref{fig2}d for a sample in air ($n=1.00$). Next, we calculate the complex reflection coefficient of the bare substrate $r_{sub}$ (Fig.~\ref{fig2}e). Finally, we retrieve the complex monolayer scattered fields $ r_{WS_{2}}$ by subtracting the substrate fields from the total fields (Fig.~\ref{fig2}f). $|r_{WS_{2}}|^{2}$ corresponds to the expected 2D layer intensity contribution in the absence of any interference with the substrate reflection. Even though the total reflection spectrum shows an approximately symmetric peak around the exciton resonance wavelength $\lambda=615$~nm (Fig.~\ref{fig2}d), the monolayer reflection is strongly asymmetrical (Fig.~\ref{fig2}f).

\begin{figure*}
\includegraphics[width=17cm]{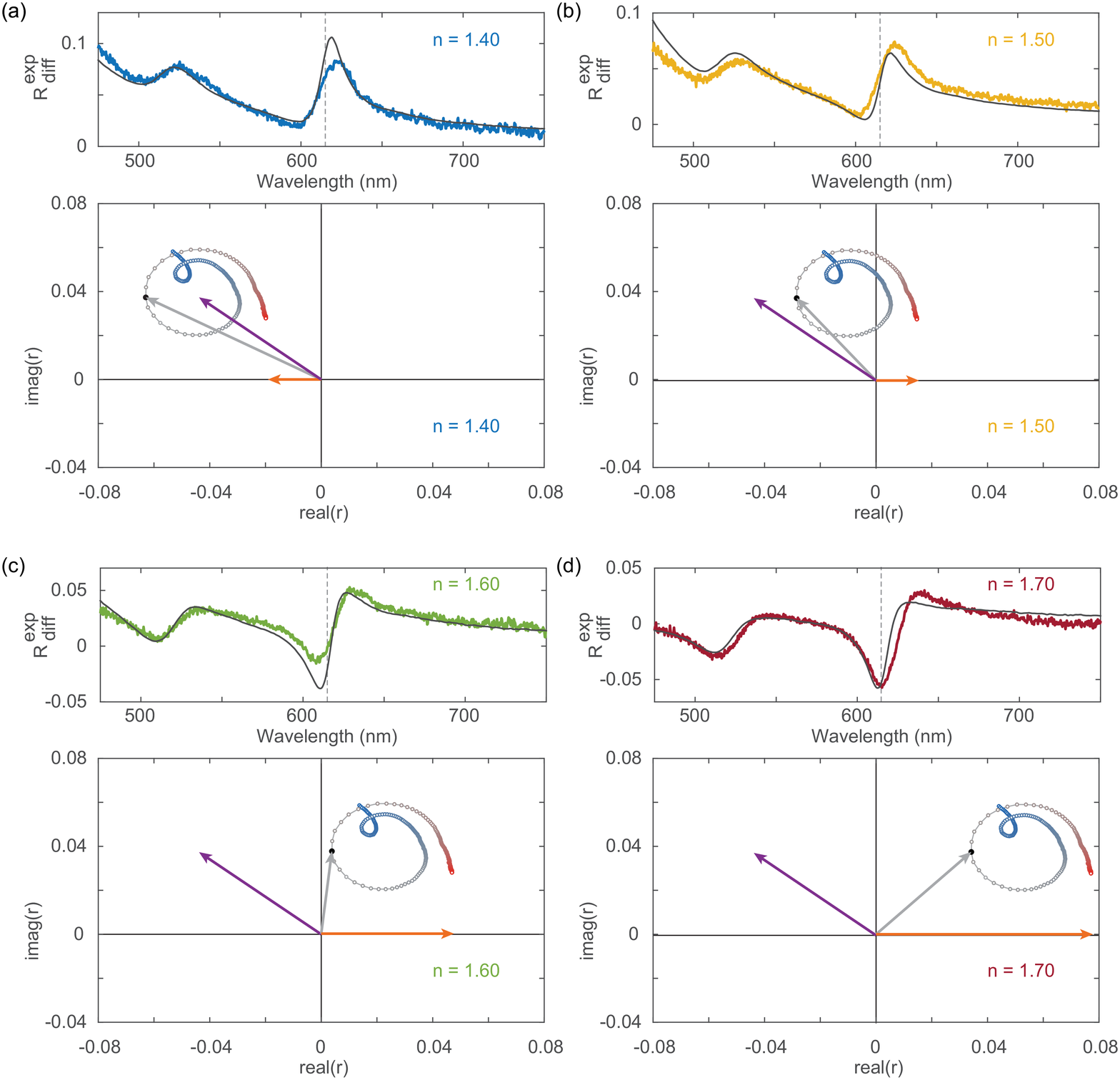}
\caption{\label{fig3}Measured (color) and calculated (black) differential reflection spectra (top) and corresponding complex phasor diagrams of the reflection coefficients (bottom) for $n=1.40$ (a), $n=1.50$ (b), $n=1.60$ (c), and $n=1.70$ (d). The orange arrow corresponds to $r_{sub}$, the purple arrow to $r_{WS_{2}}$, and the grey arrow to $r_{total}$ for $\lambda=615$~nm (indicated with black dot). The other dots show the spectral evolution of $r_{total}$ from $\lambda=475$~nm (blue dots) through $\lambda=750$~nm (red dots).}
\end{figure*}

To understand why, we now use the isolated $r_{WS_{2}}$ and $r_{sub}$ to visualize their interference by plotting the reflection amplitude and phase in the complex plane using a phasor diagram. Figure~\ref{fig3} shows the measured and calculated differential reflection spectrum (top) and the  phasor diagrams (bottom) obtained from the TMM using the experimental dielectric function values, for $n=1.40-1.70$ (a-d). The phasor diagrams show $r_{total}$ (grey) and its decomposition in the substrate reflection $r_{sub}$ (orange) and the monolayer reflection $r_{WS_{2}}$ (purple). The vectors are drawn for $\lambda=615$~nm only, while the dots show $r_{total}$ for the full spectral range ($\lambda=475-750$~nm), running from blue to red. The calculated reflection spectra are also shown (black lines).

The role of the WS$_{2}$ susceptibility (Fig.~\ref{fig2}c) in the monolayer reflection (purple) is clearly recognizable in Fig.~\ref{fig3} as the total reflection coefficient (grey) traces out the circular oscillations due to the exciton resonances. Note that the dispersion of the reflection vector is 90 degrees rotated with respect to the susceptibility in Fig.~\ref{fig2} since the fields radiated by a planar surface of radiating dipoles is delayed by this amount with respect to the driven dipole moments \cite{bohren2008absorption}. Comparing panel (a-d) shows how the substrate contribution (orange) gradually shifts from a small vector along the negative x-axis for $n=1.40$, to a larger vector along the positive x-axis for $n=1.70$. For $n=1.40$ (a) the substrate reflection is mostly in phase with the monolayer reflection, giving rise to constructive interference and thereby a peak in the total reflection around the exciton resonance wavelength. For intermediate refractive indices (e.g. $n=1.50-1.55$, b) the substrate reflection is mostly out of phase with the monolayer reflection, but the substrate reflection amplitude is small. As a result, the circular trace of $r_{total}$ intersects with $real(r)=0$, giving rise to significantly larger oscillations in the phase of the total field and a strongly asymmetrical Fano line shape. As the substrate reflection amplitude increases with the refractive index to $n=1.70$ (d), the substrate contribution is significantly larger than the monolayer reflection and the circular trace of $r_{total}$ is shifted to the right. The total reflection now oscillates towards the origin as a result of destructive interference, which gives rise to a dip in the reflection amplitude. Based on Fig.~\ref{fig3}, the (a)symmetry as well as the spectral location of the maximum of the reflection peak of the exciton line shape can thus be understood intuitively from the interference between the monolayer reflection and substrate reflection.

\begin{figure*}
\includegraphics[width=17cm]{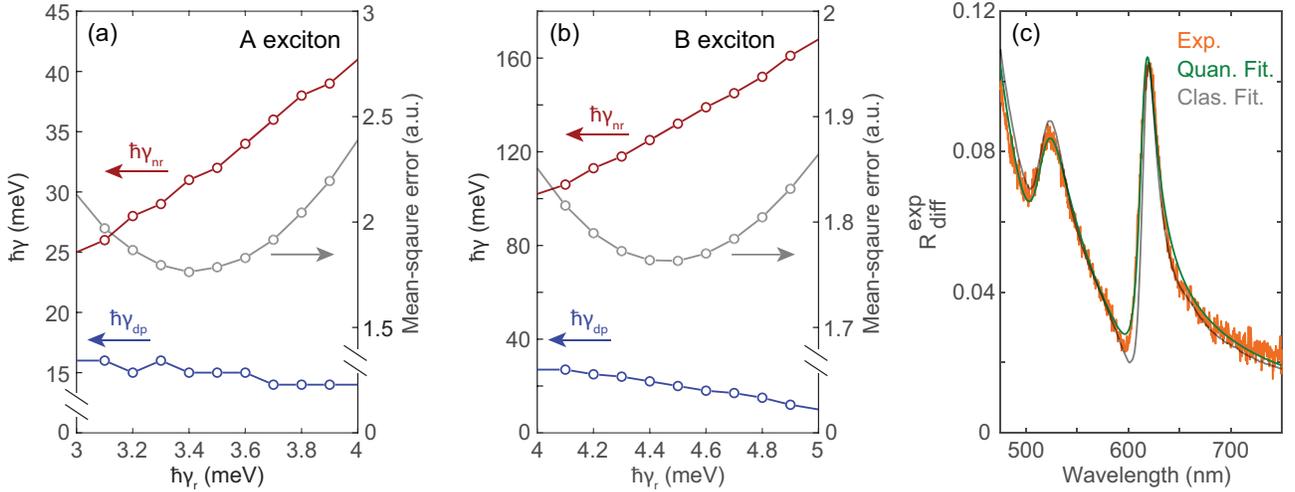}
\caption{\label{fig4}(a,b) Extracted non-radiative decay (red) and pure dephasing rates (blue) for A exciton (a) and B exciton (b) in monolayer WS$_{2}$ as a function of the presumed radiative decay rate in vacuum. The calculated mean-square error (grey) indicates the most probable intrinsic radiative decay rate of excitons in the characterized sample. (c) Measured (orange) and fitted (green for quantum model, grey for classical model) differential reflection spectra for monolayer WS$_{2}$ embedded in oil with $n=1.45$. The fitted spectra are first-principle curves based on the quantum model for the excitons, a frequency-dependent non-resonant background, and do not use ellipsometry data.}
\end{figure*}

\section{\label{sec:quantum} Quantum nature of exciton radiation} 
Sections~\ref{sec:Model}A-C evaluate the exciton line shape based on the optical constants of a thin-film WS$_{2}$ extracted from a classical ellipsometry model, and systematically examine the influence of substrate interference. However, this method is not able to capture the critical quantum and two-dimensional nature of the excitons in a WS$_{2}$ monolayer. In the quantum mechanical description developed in~\cite{Epstein2020,Scuri2018}, the excitonic emission again interferes with the light reflected from the substrate. However, the coherence can be lost due to dephasing processes. The presence of dephasing of the exciton dipole moment relative to the excitation field weakens the impact of interference effects. This directly affects the exact line shape of the excitonic features seen in the reflection spectra. To accurately describe our experimental observations, we must therefore introduce a pure dephasing rate, which is missing in the classical model. Note that we are operating at low excitation power so that we are working in the regime of perturbative quantum electrodynamics where we can work with a linear response theory and a susceptibility (as opposed to being close to saturation).

Here, we focus on a unique situation where the substrate reflection is fully suppressed ($n=1.45$ in Fig.~\ref{fig1}c) and the scattered signal purely originates from the monolayer WS$_{2}$. Under these conditions, we evaluate the influence of the quantum nature of the exciton line shape. We employ the above-mentioned quantum mechanical model to retrieve the excitonic radiative decay ($\gamma_{r}$), non-radiative decay ($\gamma_{nr}$), and pure dephasing ($\gamma_{dp}$) rates from a model fit to the asymmetric line shape observed in Fig.~\ref{fig1}c (see Supporting Information section~\ref{SI-quantum} for details). Unlike the previous works, the asymmetric line shape in our measured signal does not originate from interference between the spectrally narrow exciton radiation and broadband cavity reflection. Instead, the exciton oscillator strength of our CVD-grown WS$_{2}$ monolayer is comparable to its background dielectric constant $\varepsilon$ at room temperature, resulting into a strongly asymmetric line shape in absence of reflection of the substrate. As such, we can directly retrieve good estimates of the excitonic rates ($\gamma$) from a single reflection measurement of purely monolayer scattering.

Figure~\ref{fig4}a and b summarize the model fitting results for the A exciton and B exciton radiation in the characterized sample, respectively. We apply the mean-square error minimization fitting routine to extract the most probable non-radiative decay (red circles) and pure dephasing rates (blue circles) of excitons as a function of the presumed radiative decay rate in vacuum, which is an intrinsic parameter for monolayer WS$_{2}$. The calculated mean-square error (grey circles) indicates that the most probable intrinsic radiative decay rates for A exciton and B exciton in the characterized sample are $\hbar\gamma_{r}=3.4$~meV and $\hbar\gamma_{r}=4.5$~meV respectively, consistent with previous studies~\cite{Selig2016,Cadiz2017,Epstein2020}. Under this assumption, the A exciton exhibits a non-radiative rate of $\hbar\gamma_{nr}=31$~meV and a pure dephasing rate of $\hbar\gamma_{dp}=15$~meV, whereas B exciton shows a much larger non-radiative rate of $\hbar\gamma_{nr}=132$~meV but a similar pure dephasing rate of $\hbar\gamma_{dp}=20$~meV. This explains why the A exciton oscillator strength is much stronger than that of B exciton. A direct comparison between the quantum fitting (green) and classical fitting (grey) curves along with the measured (orange) differential reflection spectrum is shown in Fig.~\ref{fig4}c, revealing that the quantum fitting curve follows the measured spectrum more accurately. The residual mismatch between the model and the measured data around $\lambda=600$~nm in Fig.\ref{fig1}c is now resolved. The independent consideration of dephasing and non-radiative decay in the quantum model better captures the asymmetrical line shape. Despite the improved fit, low-temperature measurements of the exciton emission in the radiative limit are needed to provide conclusively measured values for $\hbar\gamma_r$. Overall, these results show that a full quantum mechanical treatment is essential to gain a full understanding of the exciton line shape, even at room temperature. 

\section{\label{sec:concl}Conclusions}
In summary, we systematically study the exciton line shape of coherently scattered light by monolayer WS$_{2}$ on a substrate. Using oils with a varying refractive index as a superstrate, we demonstrate how classical interference of the substrate reflection with the monolayer reflection can give rise to a wide range of spectral line shapes. We combine optical constants measured using spectroscopic ellipsometry with a transfer-matrix method to perform a total-field scattered-field analysis of the light reflected by the monolayer in the presence of the substrate. The analysis provides an intuitive interpretation of the role of amplitude and phase in the interference with the substrate reflection. Finally, we isolate the monolayer reflection signal experimentally using index-matching of the substrate-superstrate interface. We show that this unique configuration enables a quantum mechanical analysis to directly retrieve good estimates of the exciton dynamics without the need for substrate interference, even at room temperature. The results provide useful guidelines for the design and implementation of exciton resonances in future optical metasurfaces and nanophotonic devices.

\begin{acknowledgments}
This work was supported by a US Air Force (grant \#FA9550-17-1-0331). Some of the optical measurements were funded by the DOE "Photonics at Thermodynamic Limits" Energy Frontier Research Center under grant DE-SC0019140. JvdG was also supported by a Rubicon Fellowship and Vidi grant (VI.Vidi.203.027) from the Dutch National Science Foundation (NWO). JHS was supported by Basic Science Research Program through the National Research Foundation of Korea (NRF) funded by the Ministry of Education (NRF-2016R1A6A3A03012480). Part of this work was performed at the Nano@Stanford labs, supported by the National Science Foundation under award ECCS-1542152. 
\end{acknowledgments}


\bibliography{Droplet_references}

\clearpage
\section{\label{SI}Supporting Information}
\subsection{Ellipsometry data}
\label{SI-ellips}
The table below lists the refractive index $\tilde{n}=n+i \kappa$ measured with spectroscopic ellipsometry of a monolayer WS$_{2}$ on a fused silica substrate. The sample was obtained commercially from 2Dsemiconductors (USA).
\begin{center}
\begin{longtable}{c c c} 
 \hline
 \textbf{$\lambda$ (nm)} & \textbf{$n$} & \textbf{$\kappa$}\\
 \hline
 \endfirsthead

\hline
\textbf{$\lambda$ (nm)} & \textbf{$n$} & \textbf{$\kappa$}\\
\hline
 \endhead

 \hline
 \endfoot

 \hline
 
 \hline\hline
 \endlastfoot
451.19 & 5.55593337 & 1.60337485 \\
452.78 & 5.53207771 & 1.53539569 \\
454.38 & 5.49668218 & 1.46578764 \\
455.98 & 5.45646845 & 1.4095249 \\
457.58 & 5.4171764 & 1.34998988 \\
459.17 & 5.37484226 & 1.29829688 \\
460.77 & 5.3440806 & 1.25714654 \\
462.37 & 5.30765634 & 1.21390041 \\
463.96 & 5.27738835 & 1.1751812 \\
465.56 & 5.24297536 & 1.13836684 \\
467.16 & 5.20830196 & 1.11376767 \\
468.76 & 5.17045418 & 1.08011093 \\
470.35 & 5.14176285 & 1.05874059 \\
471.95 & 5.10666248 & 1.04117363 \\
473.55 & 5.0672122 & 1.02185808 \\
475.14 & 5.03947527 & 1.00468578 \\
476.74 & 5.00616084 & 0.992692525 \\
478.34 & 4.97701878 & 0.970653389 \\
479.93 & 4.95213215 & 0.959749187 \\
481.53 & 4.92725598 & 0.953263932 \\
483.13 & 4.89611286 & 0.940585053 \\
484.73 & 4.87309234 & 0.934541948 \\
486.32 & 4.84590475 & 0.925000225 \\
487.92 & 4.82004391 & 0.924894933 \\
489.52 & 4.7939709 & 0.927172624 \\
491.11 & 4.76246804 & 0.928320246 \\
492.71 & 4.73796834 & 0.928357502 \\
494.31 & 4.715721 & 0.938187725 \\
495.9 & 4.69640337 & 0.946143391 \\
497.5 & 4.67428708 & 0.95687095 \\
499.1 & 4.65476411 & 0.974033034 \\
500.69 & 4.63812732 & 0.997988957 \\
502.29 & 4.62738573 & 1.02030733 \\
503.88 & 4.61906726 & 1.03780127 \\
505.48 & 4.626015 & 1.07076099 \\
507.08 & 4.63300582 & 1.09948916 \\
508.67 & 4.65263819 & 1.12983829 \\
510.27 & 4.68946147 & 1.16096336 \\
511.87 & 4.73032255 & 1.18235144 \\
513.46 & 4.77107542 & 1.19923188 \\
515.06 & 4.8135575 & 1.19482027 \\
516.65 & 4.870515 & 1.18956338 \\
518.25 & 4.91600168 & 1.16207255 \\
519.85 & 4.95845672 & 1.12625691 \\
521.44 & 4.99883702 & 1.08317804 \\
523.04 & 5.02084993 & 1.02231449 \\
524.63 & 5.0407147 & 0.970977374 \\
526.23 & 5.04857399 & 0.906876269 \\
527.83 & 5.04603282 & 0.849332094 \\
529.42 & 5.04261665 & 0.794924292 \\
531.02 & 5.02480099 & 0.748356751 \\
532.61 & 5.00645214 & 0.706903937 \\
534.21 & 4.98271241 & 0.669222789 \\
535.8 & 4.95887924 & 0.637809004 \\
537.4 & 4.9237856 & 0.60584532 \\
539 & 4.90293075 & 0.59027829 \\
540.59 & 4.86839302 & 0.564435632 \\
542.19 & 4.83868339 & 0.543136216 \\
543.78 & 4.81381607 & 0.522127987 \\
545.38 & 4.78718165 & 0.512734689 \\
546.97 & 4.76284223 & 0.497091494 \\
548.57 & 4.7418898 & 0.490173328 \\
550.16 & 4.7193387 & 0.479394785 \\
551.76 & 4.69603341 & 0.47033888 \\
553.35 & 4.67170704 & 0.459096763 \\
554.95 & 4.65047879 & 0.446648316 \\
556.54 & 4.63517291 & 0.449189245 \\
558.14 & 4.61229729 & 0.440794622 \\
559.73 & 4.5891918 & 0.431883626 \\
561.33 & 4.57053876 & 0.436661819 \\
562.92 & 4.54861321 & 0.437311284 \\
564.51 & 4.53282153 & 0.43685021 \\
566.11 & 4.50866448 & 0.430757724 \\
567.7 & 4.48989959 & 0.429713676 \\
569.3 & 4.47010654 & 0.423172269 \\
570.89 & 4.44539578 & 0.418119292 \\
572.49 & 4.42029673 & 0.41795579 \\
574.08 & 4.40627306 & 0.425313613 \\
575.67 & 4.38048009 & 0.411955304 \\
577.27 & 4.3542882 & 0.410443063 \\
578.86 & 4.33122181 & 0.414823668 \\
580.46 & 4.29962886 & 0.41790129 \\
582.05 & 4.27690476 & 0.417684697 \\
583.64 & 4.24272431 & 0.426493856 \\
585.24 & 4.20656612 & 0.42764342 \\
586.83 & 4.17679339 & 0.445185295 \\
588.42 & 4.13700988 & 0.458106002 \\
590.02 & 4.09825277 & 0.477846991 \\
591.61 & 4.05230345 & 0.499154369 \\
593.2 & 4.00905027 & 0.536039539 \\
594.8 & 3.95840669 & 0.571561264 \\
596.39 & 3.91055184 & 0.632225641 \\
597.98 & 3.85638963 & 0.694114602 \\
599.57 & 3.79781206 & 0.764657098 \\
601.17 & 3.74656206 & 0.866739281 \\
602.76 & 3.72011114 & 1.01219967 \\
604.35 & 3.69568539 & 1.15978659 \\
605.95 & 3.70669801 & 1.33710655 \\
607.54 & 3.77723491 & 1.54915218 \\
609.13 & 3.92185724 & 1.76653977 \\
610.72 & 4.1443757 & 1.94732752 \\
612.31 & 4.44469903 & 2.06814679 \\
613.91 & 4.7884041 & 2.08471152 \\
615.5 & 5.12230486 & 1.98350136 \\
617.09 & 5.40062759 & 1.78423274 \\
618.68 & 5.59428867 & 1.52799834 \\
620.27 & 5.67412101 & 1.25112021 \\
621.87 & 5.68011061 & 1.00587988 \\
623.46 & 5.6266209 & 0.797572864 \\
625.05 & 5.55705916 & 0.640212842 \\
626.64 & 5.47467382 & 0.516046001 \\
628.23 & 5.38992989 & 0.428776632 \\
629.82 & 5.31348073 & 0.358665518 \\
631.41 & 5.23631533 & 0.310528941 \\
633 & 5.16813398 & 0.26736586 \\
634.59 & 5.11216149 & 0.241238782 \\
636.19 & 5.0529211 & 0.207683884 \\
637.78 & 5.00486455 & 0.203155657 \\
639.37 & 4.95828429 & 0.184214368 \\
640.96 & 4.91405042 & 0.173922673 \\
642.55 & 4.87835399 & 0.168608114 \\
644.14 & 4.84541275 & 0.156316217 \\
645.73 & 4.8194956 & 0.152418108 \\
647.32 & 4.7920169 & 0.15092899 \\
648.91 & 4.77314481 & 0.15321354 \\
650.5 & 4.75557493 & 0.146973155 \\
652.09 & 4.74229463 & 0.151530594 \\
653.68 & 4.73035402 & 0.143805006 \\
655.27 & 4.71647043 & 0.145952072 \\
656.85 & 4.70402562 & 0.138185922 \\
658.44 & 4.68951448 & 0.128937358 \\
660.03 & 4.67334395 & 0.124446118 \\
661.62 & 4.6646463 & 0.11853773 \\
663.21 & 4.64912292 & 0.106204644 \\
664.8 & 4.63080814 & 0.103270547 \\
666.39 & 4.61066979 & 0.10512476 \\
667.98 & 4.59401483 & 0.0895891282 \\
669.57 & 4.57722908 & 0.0937825716 \\
671.15 & 4.56285112 & 0.0773075594 \\
672.74 & 4.55601966 & 0.0885939924 \\
674.33 & 4.53852637 & 0.0814908583 \\
675.92 & 4.51987175 & 0.0723787896 \\
677.51 & 4.512598 & 0.0740849062 \\
679.09 & 4.48666328 & 0.0704343762 \\
680.68 & 4.47986971 & 0.0697011499 \\
682.27 & 4.47446541 & 0.0711854622 \\
683.86 & 4.46357759 & 0.0632650022 \\
685.44 & 4.45153277 & 0.0615898742 \\
687.03 & 4.44319804 & 0.069321685 \\
688.62 & 4.43155769 & 0.0688336422 \\
690.2 & 4.42737647 & 0.052075754 \\
691.79 & 4.42129603 & 0.0609916043 \\
693.38 & 4.40834495 & 0.0549402212 \\
694.96 & 4.39934593 & 0.0539670262 \\
696.55 & 4.39507832 & 0.0659113579 \\
698.14 & 4.39409351 & 0.0572225996 \\
699.72 & 4.3839524 & 0.0779073597 \\
701.31 & 4.37969284 & 0.0826261623 \\
702.89 & 4.36885751 & 0.0696932279 \\
704.48 & 4.37496221 & 0.0680176666 \\
706.07 & 4.36273278 & 0.0585823604 \\
707.65 & 4.35500222 & 0.0672004387 \\
709.24 & 4.33531098 & 0.0604048641 \\
710.82 & 4.33172561 & 0.0626470393 \\
712.41 & 4.32586575 & 0.0593886718 \\
713.99 & 4.31737111 & 0.0550712893 \\
715.58 & 4.3110822 & 0.0629833754 \\
717.16 & 4.3056154 & 0.075382921 \\
718.74 & 4.29246491 & 0.0507076368 \\
720.33 & 4.29748315 & 0.0619358548 \\
721.91 & 4.29045942 & 0.0672818484 \\
723.5 & 4.28275975 & 0.0745829286 \\
725.08 & 4.2687881 & 0.0693199744 \\
726.66 & 4.27471325 & 0.0702623569 \\
728.25 & 4.25458525 & 0.0450803744 \\
729.83 & 4.26523029 & 0.0759525599 \\
731.42 & 4.26100456 & 0.058636954 \\
733 & 4.24843685 & 0.0626718409 \\
734.58 & 4.24209 & 0.0444409697 \\
736.16 & 4.24472633 & 0.0468427839 \\
737.75 & 4.24497996 & 0.0395959473 \\
739.33 & 4.23670397 & 0.0539062471 \\
740.91 & 4.23060917 & 0.0591428999 \\
742.49 & 4.22824626 & 0.0577497336 \\
744.08 & 4.22266313 & 0.0578371417 \\
745.66 & 4.2263853 & 0.0510449763 \\
747.24 & 4.21407617 & 0.0466051473 \\
748.82 & 4.21858489 & 0.0516738643 \\
750.4 & 4.21145708 & 0.0572062887 \\
751.98 & 4.19995076 & 0.0358976356 \\
753.57 & 4.21126046 & 0.0712170556 \\
755.15 & 4.1976249 & 0.0404090207 \\
756.73 & 4.19947465 & 0.0544345765 \\
758.31 & 4.18495551 & 0.0615749611 \\
759.89 & 4.18171192 & 0.0523451376 \\
761.47 & 4.18117916 & 0.0544699465 \\
763.05 & 4.17816051 & 0.0538017351 \\
764.63 & 4.16801085 & 0.0425819797 \\
766.21 & 4.16396242 & 0.050964255 \\
767.79 & 4.1565256 & 0.0457812356 \\
769.37 & 4.15971559 & 0.0533314037 \\
770.95 & 4.16450749 & 0.0581744637 \\
772.53 & 4.14378542 & 0.0466885914 \\
774.11 & 4.14740956 & 0.0433313595 \\
775.68 & 4.14776299 & 0.0294474003 \\
777.26 & 4.12910941 & 0.0349546939 \\
778.84 & 4.14691007 & 0.0725099381 \\
780.42 & 4.15184555 & 0.072160027 \\
782 & 4.13242007 & 0.0557956558 \\
783.58 & 4.13732068 & 0.0618182419 \\
785.15 & 4.14767129 & 0.0736000196 \\
786.73 & 4.12106087 & 0.0226459148 \\
788.31 & 4.12460362 & 0.0430480252 \\
789.89 & 4.12464139 & 0.0688400633 \\
791.46 & 4.1213756 & 0.0475089935 \\
793.04 & 4.11393077 & 0.0329569587 \\
794.62 & 4.11107337 & 0.0392294595 \\
796.19 & 4.1138047 & 0.0505232043 \\
797.77 & 4.11359237 & 0.0538686964 \\
799.35 & 4.10629151 & 0.0653123974

\end{longtable}
\end{center}
\subsection{Quantum description of the reflection from WS$_2$ monolayer embedded in different dielectric environments}
\label{SI-quantum}
To calculate the reflection spectrum of the WS$_{2}$ monolayer embedded in different dielectric environments, we consider a three-layer dielectric stack, where the top, medium, and bottom layer correspond to the refractive index oil, the dielectric background of the WS$_{2}$ monolayer, and the fused silica substrate, respectively. Two planar dipole arrays are placed in the center of the medium layer ($z=0$), representing A and B excitons in the WS$_{2}$ monolayer. The positive frequency part of the field operator at position $z$ can be expressed as:
\begin{equation}
    E(z,t)=E_{0}(z,t)+\frac{k_{0}^{2}}{\varepsilon_{0}}G(z)P_{A}(t)+\frac{k_{0}^{2}}{\varepsilon_{0}}G(z)P_{B}(t)
\end{equation}
$E_{0}(z,t)$ represents the field distribution in the absence of exciton resonances, and $P_{A} (t)=\mu_{A} c_{A}(t)$ denotes the two-dimensional polarization induced by the A exciton resonance under a plane-wave illumination $E_{inc}$, where $\mu_{A}$ is the dipole moment of the exciton transition and $c_{A}(t)$ is the exciton annihilation operator. $k_{0}$ indicates the free-space wave number of the incident electric field and $\varepsilon_{0}$ is the permittivity of free space. We note that $\mu_{A}$ is fundamentally linked to the intrinsic radiative decay rate of excitons in vacuum $\gamma_{rA}$ as:
\begin{equation}
    \gamma_{rA}=\frac{k_{0} \mu_{A}^{2}}{\varepsilon_{0}}
\end{equation}
Identical relations for $P_{B}(t)$, $\mu_{B}$, $c_{B}$, and $\gamma_{rB}$ can be defined for the B exciton resonance. Meanwhile, $G(z)$ denotes the Green’s function used to connect the polarization induced by exciton resonances to its scattered electric field, and thus can be further written as $G(z)=Ge^{ikz}$ when $z$ is the position above the WS$_{2}$ monolayer. The total reflected electric field is therefore found to be:
\begin{multline}
    E_{r}(z,t)=r_{0}(z)E_{inc}+\frac{k_{0}^{2}}{\varepsilon_{0}}G(z)\mu_{A}
    c_{A}(t)\\
    +\frac{k_{0}^{2}}{\varepsilon_{0}}G(z)\mu_{B}c_{B}(t)
\end{multline}
$r_{0}(z)E_{inc}=r_{0}E_{inc}e^{ikz}$ is the reflected electric field from the non-resonant background. If we ignore the interference between the A exciton and B exciton radiation as they are spectrally misaligned, the expectation value of the reflectance at $z_{0}$ from such a material stack is given by:

\begin{multline}
\label{eq4}
R= \lvert r_{0}\rvert^{2}+\left(\frac{k_{0}^{2}\mu_{A}}{\varepsilon_{0}}\right)^{2}|G|^{2}\frac{\langle c_{A}^{\dagger} c_{A}\rangle}{|E_{inc}|^{2}}
  +\left(\frac{k_{0}^{2}\mu_{B}}{\varepsilon_{0}}\right)^{2}|G|^{2}\frac{\langle c_{B}^{\dagger} c_{B}\rangle}{|E_{inc}|^{2}}\\
  +2\frac{k_{0}^{2}\mu_{A}}{\varepsilon_{0}}\Re\left[{r_{0}^{*}G\frac{\langle c_{A} \rangle}{E_{inc}}}\right]
    +2\frac{k_{0}^{2}\mu_{B}}{\varepsilon_{0}}\Re\left[{r_{0}^{*}G\frac{\langle c_{B} \rangle}{E_{inc}}}\right]
\end{multline}
The first term indicates the non-resonant background reflection in the absence of the exciton resonance, and the second and third term are the direct reflection from the A and B exciton radiation by self-interference. The last two terms, however, evaluate the interference between the reflection from the non-resonant background and that from the A and B exciton radiation.

Up to now, the analysis seems still fully classical. Nevertheless, the quantum description becomes necessary with the presence of the dephasing of the exciton polarization relative to the excitation field, as it results in an exciton radiation contribution that is not mutually coherent with the background reflection, weakening the interference effects and thus affecting the exact line shape in the reflection spectra. For a more concise description, we focus on the A exciton here, but the conclusions hold for the B exciton as well. From Eq.~\ref{eq4}, the weakened interference due to the dephasing should lead to an inequality:
\begin{equation}
    \langle c_{A}^{\dagger} c_{A}\rangle>|\langle c_{A}\rangle|^{2},
\end{equation}
a relation that we will prove from the quantum mechanical description as discussed below. Specifically, the exciton dynamics can be described by two master equations:
\begin{equation}
    \frac{d}{dt}\langle c_{A}\rangle=i(\delta_{A}-\Delta_{A}+i\tilde{\gamma_{A}})\langle c_{A}\rangle+i \mu_{A}E_{0},
\end{equation}
\begin{equation}
     \frac{d}{dt} \langle c_{A}^{\dagger} c_{A}\rangle=
     -\gamma_{A}\langle c_{A}^{\dagger} c_{A}\rangle
     +i\mu_{A}E_{0}\langle c_{A}^{\dagger}\rangle
     -i\mu_{A}E_{0}^{\dagger}\langle c_{A}\rangle
     .
\end{equation}
In these equations, $\delta_{A}=\omega-\omega_{A}$ is the detuning of the frequency of the incident light $\omega$ with respect to the exciton resonance frequency $\omega_{A}$. $\Delta_{A}=0.5 \gamma_{rA}k_{0}\Re{\left[G(0)\right]}$ is the shift of the exciton resonant frequency modified by the local photonic environment (i.e., the Lamb shift). $\tilde{\gamma_{A}}=\gamma_{dA}+\frac{1}{2}\gamma_{A}=\gamma_{dA}+\frac{1}{2}\left(\gamma_{nrA}+2\gamma_{rA}k_{0}\Im{\left[G(0)\right]}\right)$ represents the total decay rate of excitons, being the sum of the pure dephasing rate $\gamma_{dA}$ and the population decay rate $\gamma_{A}$. The non-radiative decay rate is represented by $\gamma_{nrA}$, and $2\gamma_{rA}k_{0}\Im{\left[G(0)\right]}$ is the external decay rate modified by the local photonic environment as well (i.e., the Purcell effect). Finally, $E_{0}=E_{0}(z=0)$ for simplicity.

At steady state, the expectation value of the exciton annihilation $\langle c_{A}\rangle$ and exciton population $\langle c_{A}^{\dagger} c_{A}\rangle$ are found to be:
\begin{equation}
\langle c_{A}\rangle=
\frac{-i \mu_{A} E_{0}}
{i(\delta_{A}-\Delta_{A}+i\tilde{\gamma_{A}})},
\end{equation}

\begin{equation}
    \langle c_{A}^{\dagger} c_{A}\rangle=
    \frac{2 \mu_{A}^{2}|E_{0}|^{2}\tilde{\gamma_{A}}}
    {\gamma_{A}\left( (\delta_{A}-\Delta_{A})^{2}+\tilde{\gamma_{A}}^{2} \right)}
         .
\end{equation}
Therefore, we find that the weakened interference is quantitatively related to the presence of the dephasing, and vanishes when $\gamma_{dA}=0$:
\begin{equation}
    \langle c_{A}^{\dagger} c_{A}\rangle=\frac{2 \gamma_{dA}+\gamma_{A}}{\gamma_{A}}|\langle c_{A}\rangle|^{2}>|\langle c_{A}\rangle|^{2}.
\end{equation}
By plugging the above equations into Eq.~\ref{eq4}, we get:
\begin{multline}
\label{eq10}
R= \lvert r_{0}\rvert^{2}+k_{0}^{2}\lvert G \rvert^{2}
\frac{2\tilde{\gamma_{A}}\gamma_{rA}^{2}\lvert E_{0} \rvert^{2}}
{\gamma_{A}\left((\delta_{A}-\Delta_{A})^{2} +\tilde{\gamma_{A}}^{2} \right)\lvert E_{inc} \rvert^{2}}\\
+k_{0}^{2}\lvert G \rvert^{2}
\frac{2\tilde{\gamma_{B}}\gamma_{rB}^{2}\lvert E_{0} \rvert^{2}}
{\gamma_{B}\left((\delta_{B}-\Delta_{B})^{2} +\tilde{\gamma_{B}}^{2} \right)\lvert E_{inc} \rvert^{2}}\\
+2 \gamma_{rA}k_{0}\Re{\left[r_{0}^{*}G\frac{-i E_{0}}{i\left(\delta_{A}-\Delta_{A} +i\tilde{\gamma_{A}} \right)E_{inc})}\right]}\\
+2 \gamma_{rB}k_{0}\Re{\left[r_{0}^{*}G\frac{-i E_{0}}{i\left(\delta_{B}-\Delta_{B} +i\tilde{\gamma_{B}} \right)E_{inc})}\right]}\\
\end{multline}

We note that such an equation can be calculated analytically with a transfer-matrix approach given a set of exciton decay rates $\left(\gamma_{rA(B)}, \gamma_{nrA(B)}, \gamma_{dA(B)} \right)$. Conversely, with the measured reflection spectra in hand, we can apply a mean-square error minimization fitting routine to extract the most probable intrinsic radiative decay, non-radiative decay, and pure dephasing rates of the A and B excitons in the characterized WS$_{2}$ monolayer, as shown in Fig.~\ref{fig4}.
\end{document}